\documentclass[twocolumn,appendixfloats,twocolappendix]{aastex701}

\usepackage{natbib}
\usepackage{amsmath,amssymb}
\usepackage{xspace}
\usepackage{rotating}
\usepackage{multirow}
\usepackage{enumitem}
\usepackage{booktabs}
\usepackage{tabularx}
\usepackage{pifont}
\usepackage{bm} 

\usepackage{dcolumn}
\usepackage{macros}
\usepackage{tikz,hyperref}
\usepackage{xcolor,colortbl}
\usepackage{fdsymbol}
\usepackage{graphicx} 
\usepackage{tikz}
\usepackage{placeins}
\usetikzlibrary{arrows,positioning}
\usetikzlibrary{external}
\definecolor{mpl_blue}{HTML}{1F77B4}
\definecolor{mpl_orange}{HTML}{FF7F0E}
\definecolor{mpl_green}{HTML}{2CA02C}
\definecolor{mpl_red}{HTML}{D62728}

\usepackage[all]{hypcap}
\usepackage[caption=false]{subfig}
\newcommand{\m}[1]{\mathbf{#1}}

\def\be{\begin{equation}}
\def\ee{\end{equation}}
\newcommand{\bb}{\begin{bmatrix}}
\newcommand{\eb}{\end{bmatrix}}
\def\bea{\begin{eqnarray}}
\def\eea{\end{eqnarray}}

\def\R{\color{red}}

\defcitealias{15yrGWB}{NG15}
\defcitealias{NG}{NG12.5}

\begin{document}
\title{Multimessenger Probes of the Supermassive Black Hole Binary Population: \\The Role of Pulsar Timing Arrays}
\author{Nima Laal}
\affiliation{Department of Physics and Astronomy, Vanderbilt University, 2301 Vanderbilt Place, Nashville, TN 37235, USA}
\email{nima.laal@gmail.com}

\author{Stephen R. Taylor}
\affiliation{Department of Physics and Astronomy, Vanderbilt University, 2301 Vanderbilt Place, Nashville, TN 37235, USA}
\email{stephen.r.taylor@vanderbilt.edu}

\author{Cayenne Matt}
\affiliation{Department of Astronomy and Astrophysics, University of Michigan, Ann Arbor, MI 48109, USA}
\email{cayenne@umich.edu}

\author{Kayhan G\"{u}ltekin}
\affiliation{Department of Astronomy and Astrophysics, University of Michigan, Ann Arbor, MI 48109, USA}
\email{kayhan.gultekin@nanograv.org}
\correspondingauthor{Nima Laal}
\email{nima.laal@gmail.com}

\begin{abstract}
By inferring the gravitational wave background (GWB) from a population of supermassive black hole binaries (SMBHBs), pulsar timing arrays (PTAs) enable the study of massive black holes. In many ways, PTAs manifest the promise of a multimessenger approach to astronomy: they can constrain SMBHB population characteristics that are otherwise difficult to constrain using electromagnetic observations, such as hardening mechanisms at sub-parsec separations. In this work, we quantify this multimessenger promise using Bayesian inference of many realizations of simulated PTA data, while adopting a model for the SMBHBs that has been successfully applied to the 15-year data set of the North American Nanohertz Observatory for Gravitational Waves (NANOGrav). 
Our analyses of 200 realistic, simulated NANOGrav data sets show that there is a greater than 50\% chance of reducing the prior uncertainty in the SMBHB hardening rate by more than 50\%, and in the SMBHB evolutionary lifetime by 25--75\%. Additionally, there is an 88\% chance that PTA data can reduce the prior uncertainty in the characteristic mass variable of the galaxy stellar mass function (GSMF) by 25--50\%. For $M_{\text{BH}}$--$M_{\text{Bulge}}$ parameters (in a model without redshift evolution) and the overall normalization parameter of the GSMF, PTA data can provide only marginal information gain beyond the constraints from electromagnetic observations. Our work delineates the domains over which electromagnetic and gravitational-wave data constrain the demographics and dynamics of the supermassive black-hole binary population, offering a clearer picture of the impact of population multi-messenger astrophysics probes with PTAs.
\end{abstract}


\section{\label{sec:Introduction}Introduction}

Pulsar timing array (PTA) experiments \citep{Shazin, Detw, FosterBecker, stevebook} are a new frontier of gravitational wave (GW) astrophysics. By performing consistent, long-term, and high-precision measurements of the time of arrival (TOA) of radio pulses from an array of millisecond pulsars, PTAs build sensitivity to a variety of elusive astrophysical phenomena. Among these is the nanohertz gravitational wave background (GWB), a stochastic process resulting from the sum of individual GW emissions over a range of GW amplitudes, frequencies, sky locations, and polarizations. The nanohertz GWB is of great interest astrophysically and cosmologically, as it likely originates from a population of supermassive black hole binaries (SMBHBs) \citep{Phinney-2001,SMBHB_prime_2}, but may also have a component from processes in the early Universe \citep{15yrnewphysics}.

The PTA data used for the GWB analyses are a series of timing residuals derived from each pulsar's observed TOAs. Timing residuals are obtained by subtracting the measured TOAs from their expected values, as predicted by a timing model that accounts for pulsar-specific deterministic factors, such as spin-down rates and astrometric parameters. As a result, the timing residuals are the sum of various contributing types of stochastic processes as well as any unmodeled deterministic signal, ranging in origin from astrophysical to detector-specific \citep{NANO15_noise_budget, noise1, noise2}. 

Evidence for a GWB signal has been found to varying degrees of significance by the North American Nanohertz Observatory for Gravitational waves (NANOGrav) \citep{15yr}, the European and Indian Pulsar Timing Array (EPTA+InPTA) \citep{d2}, the Parkes Pulsar Timing Array (PPTA) \citep{d3}, the Chinese Pulsar Timing Array (CPTA) \citep{CPTA}, and MeerKAT Pulsar Timing Array \citep{Matt}. In the case of NANOGrav's 15-year data, the discovered evidence lies in the high statistical significance of distinctive correlations between the timing residuals of pairs of 67 pulsars, consistent with a statistically isotropic GWB. The reported Bayes factors in favor of these correlations range from 200 to 1000, depending on the spectral modeling choices. 

In PTAs, the fiducial GWB model is generally attributed to a stochastic process produced by a population of circular SMBHBs evolving purely due to GW emission. This is characterized by a powerlaw characteristic strain spectrum, $h_c(f)\propto f^{-2/3}$, where $f$ is the GW frequency. Notably, the amplitude of the powerlaw is not a part of the fiducial model. It is allowed to vary over a large range of values, as the range of valid amplitudes is subject to one’s model for the differential merger-rate density of the SMBHBs \citep{mag_book, 15yrastro}. For instance, to gain insight into the population and evolution of SMBHBs, \citet{SMBHB-NANO-1}, \citet{SMBHB-NANO-2}, \citet{SMBHB-NANO-3}, and \citet{SMBHB-NANO-4} map the findings of the recent PTA analyses, including the amplitude of the fiducial model, onto constraints on the orbital eccentricity of SMBHBs, environmental effects during SMBHB hardening, the galaxy stellar mass function (GSMF), and the PTA band's typical binary black hole mass, respectively. 

The motivation for our present work is to complement existing astrophysical inference efforts by examining how informative PTA data can be about the SMBHB population. We assess this by determining the extent to which PTA Bayesian inferences can update prior information about an SMBHB population model. A comprehensive study of the source of the GWB should not merely report parameter estimation results without also examining the limitations and biases inherent in the models; these define the model \textit{viability}. To interpret parameter estimation results correctly and confidently, it is essential to be aware of the viability of the models used for inferences on PTA data sets. To this end, in this paper, we devise a framework to assess the viability of a population model of SMBHBs as the source of the nanohertz GWB found in PTAs. 

Using our framework, we quantify the extent to which the degeneracies that frequently emerge in complicated SMBHB population models---especially those involving high-dimensional parameter spaces and stochastic components---complicate GW astrophysical inference. Furthermore, we identify the point at which parameter estimation reaches saturation, beyond which additional PTA data yield diminishing returns in terms of improved constraints on the model parameters. Equally critical, we assess the effect of prior probability distributions on inference results, as posterior distributions can be shaped more by prior choices than by the data itself. Collectively, the stated effects play a significant role in our ability to inform our understanding of the Universe through the GWB signal in PTA data; hence, they warrant an in-depth study.

Our work is especially relevant to multimessenger astronomy. PTA studies of SMBHBs via the measured GWB signal complement the studies of SMBHBs and galaxies based on electromagnetic observations. For example, \citet{GSMF_prior} uses data from the FourStar Galaxy Evolution Survey \citep{2016ApJ...830...51S} to constrain the GSMF. These constraints can then inform the prior probability distributions of the GWB models in PTA analyses \citep{15yrastro}, which, in turn, use PTA observations to refine and update those astrophysical constraints. However, the extent to which PTAs can further this multimessenger promise remains unknown. Using our statistical framework and software \citep{Laal_PANDORA_2025}, we quantify the sensitivity of PTAs as GW probes of the SMBHB population, thereby elucidating their role in multimessenger astronomy. We determine the parameters that PTAs can meaningfully constrain versus those that remain prior-dominated even with high-sensitivity PTA data.

The SMBHB population model to which we subject our viability tests is a semi-analytical model for SMBHB demographics and evolution that has been studied extensively in \citet{Chen2019} and \citet{15yrastro}. To aid us in achieving our goal, we use realistic and theoretical PTA simulated data \citep{Astroforcast} and the most recent advancements in PTA astrophysical inference Bayesian searches \citep{LaalAstro}. While our results are inherently model-dependent, the chosen semi-analytical SMBHB model follows standard modeling choices in the literature, making our methods and conclusions broadly applicable to both past and future PTA astrophysical inference efforts.

The paper is structured as follows. In \S\ref{sec:library}, we review the theory behind the GWB signal and our model for the population and evolution of SMBHBs. In \S\ref{sec: PTA}, we explain the details of the Bayesian inference framework used to characterize the SMBHB population that generates the GWB signal. In \S\ref{sec: sims}, we outline the details of the construction of simulated PTA data sets that we later use in \S\ref{sec: viability} to perform our study of the learnability of SMBHB population parameters using PTA data. Finally, in \S\ref{sec: Conclusion}, we present our concluding remarks and discuss the implications of our results for the future of PTAs.

\section{\label{sec:library}The GWB and a Model for SMBHB Population and Evolution}

 Throughout this work, we use natural units. Consider a general model where the sky- and polarization-averaged strain from a single binary source is measured in the source frame, $h_s$, and the total number of sources, $N$, is a function of the redshift, $z$, and a series of $p$ source parameters, collectively denoted by $\bm{\theta}$. These source parameters could include the chirp mass $M_c$ or other properties of SMBHBs that are of interest. In the observer frame, we approximate the strains as adding in quadrature to give the total observed characteristic strain of the GWB\footnote{This addition of strains in quadrature is consistent with standard SMBHB population modeling approaches in the literature, but we note that it ignores some subtleties of GW interference \citep{Lamb:2025niq}.}, $h_c$ \citep{Phinney-2001}, such that
\begin{equation}
    h_c^2(f)=\int{d\bm{\theta }\frac{dz}{1+z}\left\{\left[\frac{\partial^{p+2}N}{\partial\bm{\theta }\partial z \partial\ln{f}}\right]h_s^2(f) \right\}} \label{eq:strain general},
\end{equation}
where $f$ is the observed frequency of GWs and the $(1+z)$ factor accounts for the cosmological expansion of the Universe. The $2$ in $\partial^{p+2}$ comes from derivatives with respect to $z$ and $\ln{f}$. Based on \autoref{eq:strain general}, two quantities need to be modeled to determine the GWB's characteristic strain: (i) the strain of individual GW sources and (ii) the dependence of the number of sources on the relevant cosmological and astrophysical parameters.

The simplest description of GW emission from a compact binary system assumes an isolated, circular, Newtonian orbit \citep{Phinney-2001}, such that \citep{Kipp}
\begin{equation}
    h_{s}^{2}\left( f \right) = \frac{32}{5}\frac{M_c^{10/3}}{d_c^2} (\pi f)^{4/3} \label{eq:fiducial}.
\end{equation}
where $d_c$ denotes the comoving distance to an SMBHB at redshift $z$. 
Assuming that the binary's orbital evolution is driven only by GW radiation gives $dt/d\ln f \propto f^{-8/3}$, which, when combined with the squared strain of each source, $h_s\propto f^{4/3}$, gives a simple powerlaw relationship for the GWB strain spectrum, $h_c(f)\propto f^{-2/3}$. 

The GWB model based on \autoref{eq:fiducial} has been adopted as the fiducial model for the GWB in the PTA detection literature. However, much of what we would like to learn about SMBHBs lies in a more detailed description of the binary. For instance, one can construct models for the GWB that relax the ``isolated circular orbit'' assumption and consider effects such as dynamical friction \citep{dynamical_friction}, stellar scattering, circumbinary disk interactions \citep{15yrastro,SMBHB_prime_5,2024MNRAS.534.2609S}, and eccentricity \citep{ecc_binary_1, ecc_binary_2} to play a role in the evolution of an SMBHB system. These effects will change the frequency dependence of the GWB’s characteristic strain compared to that predicted by the fiducial model \citep{mag_book}.

In this work, we use the \textsc{phenom} SMBHB population model introduced in \citet{15yrastro} and implemented in the software \texttt{holodeck} \citep{holodeck}. \textsc{Phenom} assumes circular binaries and the Newtonian approximation, but allows for environmentally driven SMBHB hardening models. The motivation for using \textsc{phenom} is its extensive use in prior works \citep{15yrastro, Chen2019} and its simplicity. \textsc{Phenom} summarizes the rich physics of SMBHBs in terms of only a few key astrophysical parameters, making it easier to study using our statistical analysis framework of \S\ref{sec: PTA}.

More explicitly, the free parameter $\nu_{\text{inner}}$ of \textsc{phenom} controls the hardening rate of binaries as they approach and enter the regime where the frequency of their GW emission is observable by PTAs:
\begin{equation}
    \frac{db}{dt}={{H}_{b}}\,{{\left( \frac{b}{{{b}_{c}}} \right)}^{1-{{\nu }_{\text{inner}}}}}{{\left( 1+\frac{b}{{{b}_{c}}} \right)}^{{{\nu }_{\text{inner}}}-{{\nu }_{\text{outer}}}}} \label{sem-a}.
\end{equation}
In the above, $b$ is the semi-major axis of the binary, and $H_b$ is a normalization factor related to the total lifetime of the binary. The critical separation, $b_c$, is a reference value such that if $b\gg {{b}_{c}}$, the evolution of the binary is dominated by the $\nu_{\text{outer}}$ term (i.e., large separations) and if $b\ll {{b}_{c}}$, the evolution is dominated by the $\nu_{\text{inner}}$ term (i.e., small separations). We fix $\nu_{\text{outer}}$ to 2.5, as this is the expected value to use to account for dynamical friction impacting SMBHBs that end up having approximately circular orbits in the PTA band \citep{10.1093/mnras/stw2452}.
Additionally, the total lifetime of the binary, measured from the time it is formed to the time it reaches its innermost stable circular orbit (ISCO), is parameterized by $\tau_f$ in units of Gyr; it plays a crucial role in describing the binary's evolution. More explicitly, $\tau_f$ enters our model by determining the numerical value of $H_b$ in \autoref{sem-a} via
\begin{equation}
    {{\tau }_{f}}=\int_{{{b}_{\text{init}}}}^{{{b}_{\text{isco}}}}{{{\left( \frac{db}{dt} \right)}^{-1}}db}.
\end{equation}
Solving for $H_b$ leads to
\begin{equation}
    {{H}_{b}}=\tau _{f}^{-1}\int_{{{b}_{\text{init}}}}^{{{b}_{\text{isco}}}}{\left( {{\left( \frac{b}{{{b}_{c}}} \right)}^{{{\nu}_{\text{inner}}}-1}}{{\left( 1+\frac{b}{{{b}_{c}}} \right)}^{{{\nu}_{\text{outer}}}-{{\nu}_{\text{inner}}}}} \right)db}.
\end{equation}

In practice, the term involving the total number of sources in \autoref{eq:strain general} is first calculated in terms of observable properties of galaxy pairs. Furthermore, a one-to-one correspondence between galaxy pairs and SMBHBs is assumed, and an SMBHB–host relationship to translate from galaxies to SMBHBs is adopted. For \textsc{phenom}, the relevant galaxy pair properties are the number density of galaxy mergers $\eta$, stellar mass of the primary galaxy $M_{\star 1}$, and the stellar mass ratio $q_{\star}$. The properties of SMBHBs considered are the total binary mass $M$ and the binary mass ratio $q$. To express these mathematically, let,
\begin{equation}
n=\frac{dN}{dV_c},
\end{equation}
be the number density of the sources per unit co-volume $V_c$. Subsequently,
\begin{equation}
    \frac{{{\partial }^{4}}N}{\partial M\partial q\partial z\partial \ln \left( f \right)}=\frac{{{\partial }^{3}}n}{\partial M\partial q\partial z}\frac{\partial t}{\partial \ln \left( f \right)}\frac{\partial z}{\partial t}\frac{\partial {{V}_{c}}}{\partial z},
\end{equation}
for $\frac{\partial z}{\partial t}\frac{\partial {{V}_{c}}}{\partial z}=4\pi \left( 1+z \right)d_{c}^{2}$  \citep{Hogg} where $d_c$ is the co-moving distance and $\frac{\partial t}{\partial \ln(f)}$ is the amount of time that an SMBHB spends emitting at a logarithmic frequency $\ln(f)$ based on the details of the hardening model. Moreover, assuming a one-to-one correspondence between galaxy pairs and SMBHBs, one can write
\begin{align}
    {\frac{\partial^3 n}{\partial M \, \partial q \, \partial z} = 
        \frac{ \partial^3 \eta}{\partial M_{\star 1} \, \partial q_\star \, \partial z} \frac{\partial M_{\star 1}}{\partial M} \frac{\partial q_\star}{\partial q}.} \label{eq:pop-synth}
\end{align}
The term $\frac{ \partial^3 \eta}{\partial M_{\star1} \, \partial q_\star \, \partial z}$ requires the introduction of three functions:
\begin{equation}
    \frac{\partial^3 \eta }{\partial {{M}_{\star1}}\partial {{q}_{\star}}\partial z}=\frac{\Psi \left( {{M}_{\star1}},{z}' \right)}{{{M}_{\star1}}\ln \left( 10 \right)}\frac{P\left( {{M}_{\star1}},{{q}_{\star}},{z}' \right)}{{{T}_{\text{gal-gal}}}\left( {{M}_{\star1}},{{q}_{\star}},{z}' \right)}\frac{\partial t}{\partial {z}'}.
\end{equation}
The galaxy stellar mass function (GSMF) $\Psi$ is the differential number density of galaxies per decade of stellar mass, the galaxy pairing fraction $P$ is the fraction of galaxies that have a close companion galaxy, and the galaxy merger time $T_{\text{gal-gal}}$ is the time it takes for two galaxies that are gravitationally bound to complete a merger. For \textsc{phenom}, the GSMF has two free parameters $\psi_0$ and $m_{\psi,0}$, and follows the relationship
\begin{equation}
     \Psi =\ln \left( 10 \right){{\Psi }_{0}}{{\left[ \frac{{{M}_{\star1}}}{{{M}_{\psi }}} \right]}^{{{\alpha }_{\psi }}}}{{\text{exp}}{\left( -\frac{{{M}_{\star1}}}{{{M}_{\psi }}} \right)}}  
\end{equation}
where
\begin{align}
    \begin{split}
  {{\log }_{10}}\left( \frac{{{\Psi }_{0}}}{\text{Mp}{{\text{c}}^{-3}}} \right)&={{\psi }_{0}}+{{\psi }_{z}}z, \\ 
  {{\log }_{10}}\left( \frac{{{M}_{\psi }}}{{{\text{M}}_{\odot }}} \right)&={{m}_{\psi ,0}}+{{m}_{\psi ,z}}z, \\ 
  {{\alpha }_{\psi }}&=1+{{\alpha }_{\psi ,0}}+{{\alpha }_{\psi ,z}}z.
    \end{split}
\end{align}
Here, $\psi_0$ is the normalization of the GSMF, shifting the value of $\Psi_0$ up and down in the log-space. Similarly, $m_{\psi,0}$ is the characteristic mass normalization of the GSMF at $z=0$. Both parameters vary in base-10 logarithmic space. For the remaining parameters, we use $\psi_z=-0.6$, $m_{\psi,z}=0.11$, $\alpha_{\psi,0}=-1.21$, and $\alpha_{\psi,z}=-0.03$ based on \citet{Chen2019,GSMF_prior}. Lastly, the other two functions are set to their observational values based on \citet{Conselice2003,Bluck2012,Mundy2017,Duncan2019} for galaxy pair fraction and \citet{Conselice2008,BoylanKolchin2008,Conselice2009,Snyder2017} for galaxy merger time.

Finally, we consider an $M_{\text{BH}}$--$M_{\text{Bulge}}$ relationship of the form \citep{mmbulg}
\begin{equation}
    \log_{10} \left( \frac{M_{\rm BH}}{M_\odot} \right) =  \alpha \log_{10} \left( \frac{M_{\rm bulge}}{10^{11} M_\odot} \right) + X,
\end{equation}
where $X$ is a random number drawn from a Normal distribution with mean $\mu$ and variance $\epsilon_\mu^2$, $M_{\rm BH}$ is the mass of each component of the binary, $\alpha=1.1$ (average of findings from \citealt{kormandy} and \citealt{McConnell}) is a constant, and $M_{\rm bulge}$ is the stellar bulge mass. The free parameters $\mu$ and $\epsilon_\mu$ describe the dimensionless black hole mass normalization and the intrinsic scatter, respectively, both of which vary in \textsc{phenom}.

 \section{\label{sec: PTA}Bayesian Inference using PTA Experiments}
 The apparent redshift (or blueshift) to the time of arrival of pulses emitted from pulsars due to GWs is \citep{hd83, Detw, mag_book}
\begin{equation}
    z_I(t, \hat{\Omega}) = \frac{1}{2} \frac{\hat{r}^l \hat{r}^m}{1 + \hat{\Omega} \cdot \hat{r}} \Delta h_{lm}(t, \hat{\Omega}), \label{eq:red-shift}
\end{equation}
where $\hat{r}$ is the unit vector pointing in the direction of the pulsar, $\hat{\Omega}$ is the unit propagation direction of GW, $l$ and $m$ are indices representing the spatial coordinates, the index $I$ denotes a pulsar, and $t$ is the observing time. The term $\Delta h_{lm}$ represents the difference between the values of the GW metric perturbation at Earth and at the pulsar emission times. The integral of \autoref{eq:red-shift} evaluated from starting time $t_0$ to observation time $t$ yields the timing residual $\bm{\delta t_I}$:
\begin{equation}
    \bm{\delta t_I}=\int_{{{t}_{0}}}^{t}{z_I\left( {{t}'} \right)d{t}'},
\end{equation}
which is the observed quantity in PTA GW searches.

The majority of Bayesian GW detection algorithms applied to PTA data are intended for the frequency domain \citep{codereview, Laal:2023etp, OPTSTAT0}.
Even though the observing times are highly irregular in PTAs because of varying timing cadences and gaps in the data, a sparse, regularly-spaced subset of observing times can be selected for performing a Fourier transform to find the Fourier coefficients that represent the GW signal in the frequency domain \citep{AidenNeil}. For a GWB made from a finite number of sources, the Fourier coefficients from individual sources could be added together to form the Fourier coefficients of the GWB if the sources share a common set of frequencies.

In practice, we can make several approximations based on the specifics of our GWB model, the most notable of which is the assumption of Gaussianity. For instance, for a statistically isotropic, unpolarized, stationary GWB made from a sufficiently large number of independent GW sources, the GWB Fourier coefficients can be drawn probabilistically from a zero-mean multivariate Gaussian distribution with an unknown PSD covariance matrix. In this work, we adopt a simple GWB model that satisfies this Gaussianity assumption and is consistent with previous efforts \citep{15yrastro}.

Let $\bm{a}$ (units of seconds) denote the set of Fourier coefficients of all pulsars. These coefficients characterize the discrete Fourier transform of the GWB's contribution to the timing residual of all pulsars at all frequency bins. $\bm{a}$ is a matrix of size $2n_f\times n_p$, where $n_f$ is the number of GWB frequency bins considered, and $n_p$ s the number of pulsars in the PTA. In this Fourier representation, there are two coefficients for pulsar $I$ at frequency bin $k$: $a_{I;k}^{\sin}$ and $a_{I;k}^{\cos}$. Additionally, let $\bm{\rho}^2$ (units of $\text{seconds}^2$) be a set of parameters describing the amplitude of the PSD of the GWB-induced timing residuals; $\bm{\rho}^2$ is a vector of size $n_f$, representing the variance of $\bm{a}$, such that
\begin{equation}
    \left\langle a_{I;k}^{\sin }a_{J;k}^{\sin } \right\rangle =\rho _{k}^{2}{{\Gamma }_{IJ}}=\left\langle a_{I;k}^{\cos }a_{J;k}^{\cos } \right\rangle \label{cross-terms}.
\end{equation}
where $\left\langle \cdots  \right\rangle$ denotes the ensemble average over many realizations of the GWB.
The cross-correlation PSD, $\rho _{k}^{2}{{\Gamma }_{IJ}}$\footnote{For the case where $I=J$, $\Gamma$ is set to 1.}, uses the Hellings and Downs \citep{hd83} correlation values $\Gamma_{IJ}$. Finally, let $\bm{\theta}$ denote the set of SMBHB population and evolution parameters introduced in \S\ref{sec:library} which influence the timing residual PSD. In this work, $\bm{\theta}$ is a vector of length six since \textsc{phenom} has only six free parameters. 

To build the appropriate Bayesian inference framework, the above three sets of the GWB parameters are assigned prior probabilities in the recursive conditioning form
\begin{equation}
    \pi \left( \bm{\rho} ,\bm{\theta} ,\bm{a} \right)=\pi \left( \left. \bm{a} \right|\bm{\rho} ,\bm{\theta}  \right)\pi \left( \left.\bm{ \rho}  \right|\bm{\theta}  \right)\pi \left( \bm{\theta}  \right),
\end{equation}
where $\pi$ is used to emphasize the nature of the probability densities being prior distributions. Since there is no explicit $\bm{\theta}$ dependence in the way Fourier coefficients are modeled\footnote{This assumption will not hold for a GWB modeling where the impact of finite sources/anisotropy is of interest.}, one can simplify
\begin{equation}
    \pi \left( \left. \bm{a} \right|\bm{\rho} ,\bm{\theta}  \right)=\pi \left( \left. \bm{a} \right|\bm{\rho}  \right).
\end{equation}
As discussed earlier, in our model of the GWB given by \autoref{cross-terms}, $\pi \left( \left.\bm{a}  \right|\bm{\rho}  \right)$ is assumed to be a zero-mean multivariate Gaussian distribution with an unknown covariance matrix $\varphi$. Moreover, the correlation between frequency bins is set to zero. Note that the Gaussian nature of $\pi \left( \left.\bm{a}  \right|\bm{\rho}  \right)$ as well as the use of independent frequency bins are simply approximations valid in the limit of many independent sources of the GWB with circular orbits and long timing baselines \citep{JoeNeil,MVSK,Lamb:2025niq,1999PhRvD..59j2001A}. For consistency with previous work on astrophysical inference using PTAs, we make the same assumptions, treating the GWB as a stochastic noise process composed of GWs from a sufficiently large number of independent, isotropic sources. The covariance matrix $\varphi$ is formed using \autoref{cross-terms} such that
\begin{equation}
    {{\varphi }_{IJ;k}}=\rho _{k}^{2}{{\Gamma }_{IJ}},
\end{equation}
or more explicitly,
\begin{equation}
   \begin{aligned}
   \pi \left( \left. \bm{a} \right|\bm{\rho}  \right)&=\prod_{k=1}^{2n_f}\text{Normal}\left( 0,\varphi_k  \right) \label{a_gwb_draw}.
\end{aligned}
\end{equation}

The prior $\pi \left( \bm{\theta}  \right)$ is the same joint distribution used by \texttt{holodeck} to define the free parameters of \textsc{phenom} and our prior uncertainty about their values. This distribution could take any form. However, we choose to rely on astrophysical observations (see \S\ref{sec:library}) to specify four out of the six parameters of $\bm{\theta}$ to be drawn from a single-variate normal distribution. Assuming the parameters are separable,
\begin{equation}
    \pi \left( \bm{\theta}  \right) = \pi \left( {{\psi }_{0}} \right)\pi \left( {{m}_{\psi ,0}} \right)\pi \left( \mu  \right)\pi \left( {{\varepsilon }_{\mu }} \right)\pi \left( {{\tau }_{f}} \right)\pi \left( {{\nu }_{\text{inner}}} \right),
\end{equation}
with
\begin{align}
\begin{split}
   \pi \left( {{\tau }_{f}} \right)&=\text{Uniform}\left( 0.1,11 \right) \text{Gyr}, \\ 
  \pi \left( {{\psi }_{0}} \right)&=\text{Normal}\left( -2.56,0.4 \right), \\ 
  \pi \left( {{m}_{\psi ,0}} \right)&=\text{Normal}\left( 10.9,0.4 \right), \\ 
  \pi \left( \mu  \right)&=\text{Normal}\left( 8.6,0.2 \right), \\ 
  \pi \left( {{\varepsilon }_{\mu }} \right)&=\text{Normal}\left( 0.32,0.15 \right), \\ 
  \pi \left( {{\nu }_{\text{inner}}} \right)&=\text{Uniform}\left( -1.5,0 \right).
  \end{split}
  \label{astro-prior}
\end{align}

\begin{table*}[ht]
\centering
\captionsetup{font=small}
\label{table:syms}
\renewcommand{\arraystretch}{1.2}
\footnotesize

\begin{minipage}[t]{0.48\textwidth}
\centering
\begin{tabular}{c p{6.2cm}}
\toprule
\textbf{Symbol} & \textbf{Meaning} \\
\midrule
\(\displaystyle n_p \) & The number of pulsars\\
\(\displaystyle n_f \) & The number of frequency bins \\
\(\displaystyle \bm{\delta t_I} \) & Timing residuals for pulsar \(I\) in seconds. A vector of length \(M\), where \(M\) is the number of observations. \\
\(\displaystyle \bm{a} \) & Fourier coefficients across all pulsars and frequencies. Matrix of shape \(2n_f \times n_p\), units of seconds. \\
\(\displaystyle \bm{\rho} \) & Amplitudes of the PSD of the GWB-induced timing residuals. Vector of length \(2n_f\), units of seconds. \\
\(\displaystyle \bm{\theta} \) & Vector of SMBHB population and evolution parameters. Length 6 in \textsc{phenom}. \\
\bottomrule
\end{tabular}
\end{minipage}
\hfill
\begin{minipage}[t]{0.48\textwidth}
\centering
\begin{tabular}{c p{6.2cm}}
\toprule
\textbf{Symbol} & \textbf{Meaning} \\
\midrule
\(\displaystyle \tau_f \) & Controls total lifetime of SMBHB population, in Gyr. \\
\(\displaystyle \nu_{\text{inner}} \) & Controls hardening rate; dimensionless index. \\
\(\displaystyle \psi_0, m_{\psi0} \) & Two dimensionless parameters controlling the stellar mass function. \\
\(\displaystyle \mu, \epsilon_\mu \) & Dimensionless black hole mass normalization and intrinsic scatter. \\
\textsc{simGWB} & Simulated optimistic PTA data set with 90 well-timed pulsars and 10 nanoseconds timing over 20 years. \\
\textsc{A4Cast} & Realistic PTA data set based on 45 pulsars from \citet{Astroforcast} using NANOGrav 12.5-year data. \\
\bottomrule
\end{tabular}
\end{minipage}
\caption{\small Important quantities/terms involved in inference, their mathematical symbols, and a brief description of each.}
\end{table*}

Since \texttt{holodeck} generates samples drawn from $\pi \left( \left.\bm{\rho}  \right|\bm{\theta}  \right)$---but does not provide the probability distribution itself---these samples must be converted into multivariate probability density estimators before they can be used within our Bayesian inference framework. The dimensionality of the density estimation problem (six parameters within $\bm{\theta}$ in addition to $n_f$ parameters in $\bm{\rho}$) can make this challenging in a way that retains fidelity with respect to the original underlying distributions. In \citet{LaalAstro}, we demonstrate how \textit{normalizing flows} can overcome these challenges and accurately emulate $\pi \left( \left.\bm{\rho}  \right|\bm{a}  \right)$ after being trained on a library of generated samples from \texttt{holodeck}.

The PTA data likelihood, $\mathcal{L}$, is yet another multivariate Gaussian distribution, as the white noise of PTAs is Gaussian. The likelihood is factorized over the likelihoods of individual pulsars, such that
\begin{equation}
\begin{aligned}
   \mathcal{L}\left( \left. \bm{\delta t} \right| \bm{a}, \bm{a}_{\text{non-GWB}}\right)&=\prod\limits_{I=1}^{{{n}_{p}}}{   \text{Normal}\left( \bm{\mu_I} ,D_I \right)}, \\ 
  \bm{\mu_I} &={{F_I}}{{\bm{a_I}}}+{{F}_{I;\text{non-GWB}}}{{\bm{a}}_{I;\text{non-GWB}}}. 
\end{aligned}
\label{eq:likelihood}
\end{equation}
In the above, ``non-GWB'' refers to any non-GWB red noise, and the basis matrix $F$, responsible for performing the Fourier transform of the coefficients, is defined such that 
\begin{equation}
    F^\mathrm{T} = 
        \begin{bmatrix}
            \sin(2\pi f_1 t_1) & \sin(2\pi f_1 t_2) & \cdots & \sin(2\pi f_1 t_M) \\
            \cos(2\pi f_1 t_1) & \cos(2\pi f_1 t_2) & \cdots & \cos(2\pi f_1 t_M) \\
            \vdots & \vdots & \ddots & \vdots \\
            \sin(2\pi f_{n_f} t_1) & \sin(2\pi f_{n_f} t_2) & \cdots & \sin(2\pi f_{n_f} t_M) \\
            \cos(2\pi f_{n_f} t_1) & \cos(2\pi f_{n_f} t_2) & \cdots & \cos(2\pi f_{n_f} t_M)
        \end{bmatrix},
\label{eq:FT}
\end{equation}
where index $M$ ranges to the last observed TOA of a pulsar \citep{10.1093/mnras/stv1538}. Despite sharing TOAs, $F$ and $F_{\text{non-GWB}}$ do not need to share frequencies. The matrix $D$ is the timing-model-marginalized white-noise covariance matrix. In most GWB searches, $D$ is determined before the analysis begins by studying the noise profile of individual pulsars and finding the maximum-likelihood values of the white noise parameters that shape $D$, while marginalizing over uncertainties in the timing model; more details can be found in \citet{codereview}.

For real PTA data sets, as well as the simulations used later in \S\ref{sec: viability}, individual pulsars exhibit their own non-GWB noise. Non-GWB noise could originate from sources such as the interstellar medium or solar wind \citep{noise2,2024ApJ...972...49L}. The inclusion of these non-GWB parameters in our Bayesian analysis scheme is trivial, as such parameters do not introduce correlations between frequency bins or pulsars. They simply model any non-GWB noise on a pulsar-by-pulsar basis. However, they increase the difficulty of Markov Chain Monte Carlo (MCMC) sampling of the model posterior distribution by expanding the parameter space and introducing significant degeneracies with the GWB and other GW signals.

For the sake of consistency with previous work, and to simplify the noise modeling as much as possible, the non-GWB noise is modeled as a powerlaw with varying amplitude and spectral index; this is akin to each pulsar having achromatic spin noise. More specifically, each pulsar $I$ has a non-GWB contribution to the PSD of timing residuals, ${\rho^{2}_{I;k\text{;non-GWB}}}$, parametrized by
\begin{equation}
    {\rho^{2}_{I;k\text{;non-GWB}}}=\frac{A_{I}^{2}}{12{{\pi }^{2}T}{{f_k}^{3}}}{{\left( \frac{f_k}{{{f}_{\text{yr}}}} \right)}^{3-{{\gamma }_{I}}}} \label{eq:powerlaw_I},
\end{equation}
for $f_\text{yr}=1/1\,\text{yr}$, and $T$ being the observation baseline of the PTA. For the prior probabilities, $\log_{10}{A_I}$ is assigned a uniform distribution between $-18$ and $-11$, while $\gamma_I$ is assigned a uniform prior between $0$ and $7$. Lastly, for $\bm{a}_{I;\text{non-GWB}}$, the Gaussian prior 
\begin{equation}
    \pi \left( {{a}_{I;\text{non-GWB}}} \right)=\prod\limits_{k=1}^{2{{n}_{f}}}{\text{Normal}\left( 0,{{\rho^2 }_{I;k;\text{non-GWB}}} \right)}
\end{equation}
is chosen to be consistent with previous efforts \citep{15yrastro,codereview}.

After accounting for non-GWB noise, the Bayesian framework is complete and can be connected to an MCMC algorithm to sample from the joint parameter posterior distribution. Refer to \autoref{table:syms} for a glance at the important quantities and terms of the GWB inference discussed in this work.

 \section{\label{sec: sims}Simulations}
 \begin{figure}[htbp]
\includegraphics[width=\linewidth]{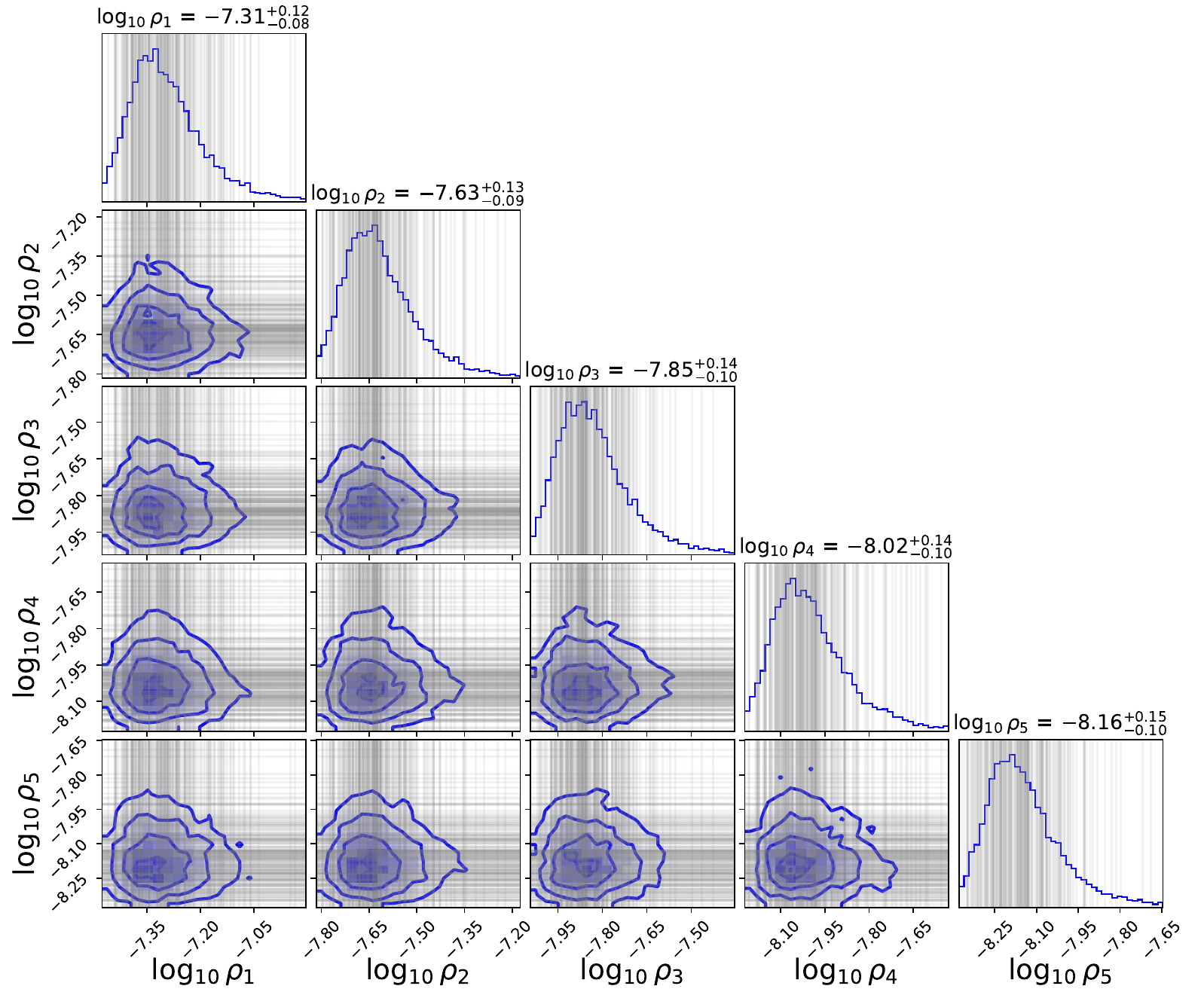}
    \caption{A set of contours and histograms showing the multivariate distribution of the GWB-induced PSD of the timing residuals that is used for simulations in this work. The solid blue contours highlight the distribution, while the gray lines indicate the 200 samples drawn from it, which are used to inject into the simulated data sets \textsc{A4cast} and \textsc{simGWB}. As shown in the figure, the 200 drawn samples cover most of the distribution.}
    \label{sim_pos}
\end{figure}

 \begin{figure*}[htbp]
\includegraphics[width=\linewidth]{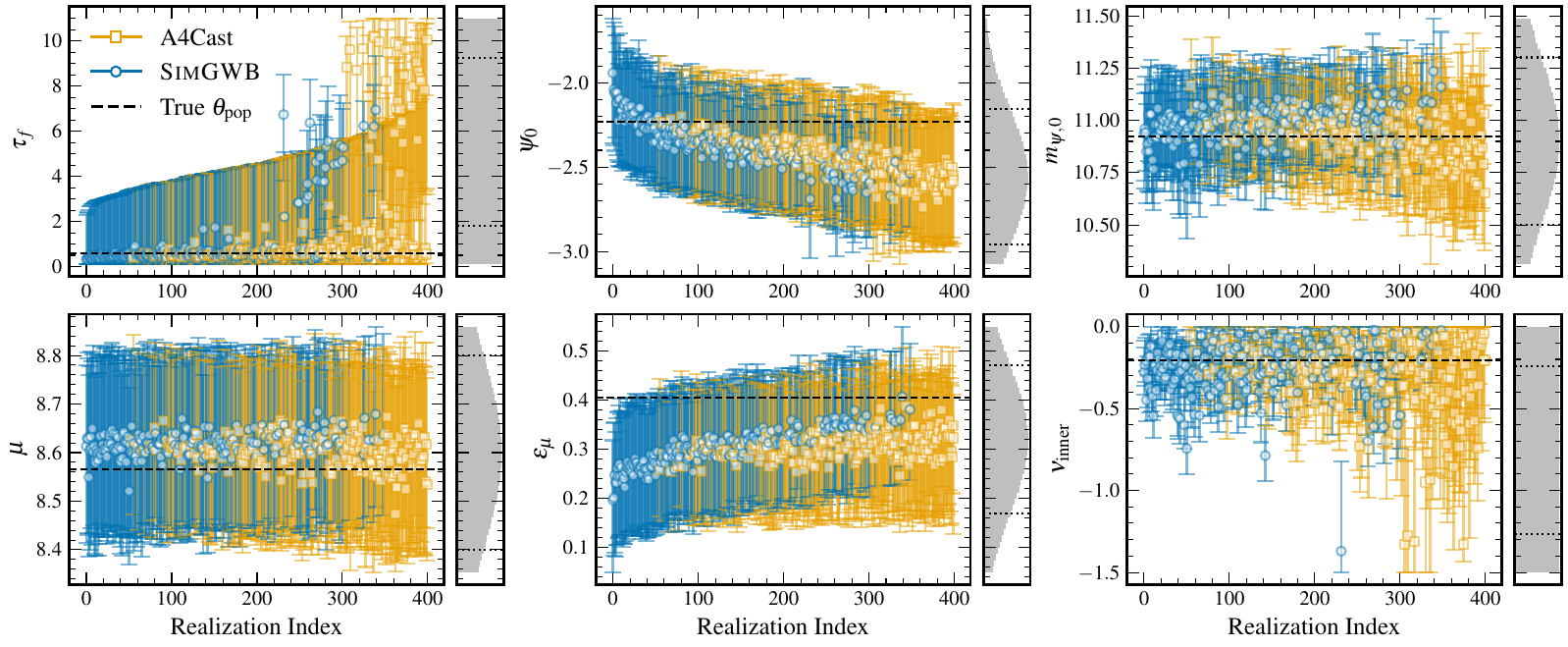}
    \caption{A set of plots depicting the Bayesian inference posterior distributions for $\bm{\theta}$ obtained by analyzing all realizations of \textsc{A4cast} (orange squares) and \textsc{SimGWB} (blue circles). The error bars range over the 68\% HDI, whereas the circles and squares show the MAP values. Each panel of the figure belongs to one parameter in $\bm{\theta}$. Attached to each panel, the prior probability distribution and its 68\% HDI interval (dashed horizontal lines) are shown. The actual values of each parameter used in all simulations are shown with a dashed line across the main panels. The main takeaway messages of the plots are (i) the prior-dominant posterior distributions for the $M_{\text{BH}}$--$M_{\text{Bulge}}$ parameters regardless of the data set and (ii) informative posterior distributions for the parameters that cannot be easily constrained using electromagnetic observations (i.e., $\tau_f$ and $\nu_{\rm{inner}}$).}
    \label{postcomp}
\end{figure*}

In this work, we use the \textsc{A4cast} \citep{Astroforcast} simulation framework to specify the pulsar-level parameters. These parameters set how long each pulsar is observed for, how irregular the consequent observation events are, how much uncertainty is in the observation of individual TOAs (i.e., the white noise level), and what spectral properties ($A_I$ and $\gamma_I$ of \autoref{eq:powerlaw_I}) the non-GWB noise possesses. \textsc{A4cast} is based on the NANOGrav 12.5-year data set \citep{12year}, comprising 45 pulsars, but is extrapolated to extend its overall observational baseline to 20 years.

To isolate the complications introduced by the GWB noise from other factors, and to create an idealized PTA data set, we also create a second type of simulated data, which we refer to as \textsc{SimGWB}. \textsc{SimGWB} considers 90 pulsars distributed statistically uniformly in the sky that are identical to each other in every aspect except for their position on the sky. All pulsars are timed for 20 years with a 14-day regular cadence while having a negligible level of white noise (10~ns) and no non-GWB red noise. 

To specify the global properties of our simulated PTA, including the exact frequency bins, the amplitudes of the GWB-induced timing residual PSD at the selected frequency bins, and the underlying set of SMBHB evolution parameters $\bm{\theta}$, we make the following well-motivated choices:
\begin{itemize}[leftmargin=*]
    \item[]\textbf{Frequency bins---} To reduce the computational cost of the Bayesian inferences, we only analyze on the lowest five Fourier sampling frequencies: $1/(20\,\rm{yrs})$, $2/(20\,\rm{yrs})$, $3/(20\,\rm{yrs})$, $4/(20\,\rm{yrs})$, and $5/(20\,\rm{yrs})$.
    
    \item[]\textbf{SMBHB evolution parameters---} The chosen $\bm{\theta}$ is
    \begin{equation}
\begin{aligned}
    \tau_f            &= 0.594 \text{~Gyr},           & \psi_0          &= -2.23, 
    & m_{\psi ,0}       &= 10.92, \\             \mu             &= 8.56,
    & \varepsilon_{\mu} &= 0.406,          & \nu_{\text{inner}} &= -0.205. 
\end{aligned}
\label{astro-chosen}
\end{equation}      
This choice is motivated by the fact that the four parameters following univariate normal distributions in \autoref{astro-prior} are set to values near the medians of their respective prior distributions. As a result, the normalizing flow emulator receives a sufficient number of training samples to accurately learn the mapping between $\bm{\theta}$ and $\bm{\rho}$. Additionally, such choices imply that our simulated data sets are consistent with current astrophysical expectations.

\item[]\textbf{PSD of the GWB-induced timing residuals---} We model the number of SMBHBs in a given realization of the Universe probabilistically by drawing samples from a Poisson distribution. In \textsc{phenom}, \autoref{eq:strain general} is implemented as
\begin{align} \label{pois}
\begin{split}
    h_{c}^{2}=\sum\limits_{z,M,q,f} & \mathcal{P}\left( \frac{{{\partial }^{4}}n}{\partial M\partial q\partial z\partial \ln \left( f \right)}\Delta M\Delta q \frac{\Delta z}{1+z}\Delta \ln \left( f \right) \right) \\ 
    & \times \text{ }\frac{h_{s}^{2}\left( f; z,M,q \right)}{\Delta \ln \left( f \right)},
\end{split}
\end{align}
where $\Delta$ denotes finite bins of a given parameter, and $h_c^2$ is later converted to $\bm{\rho}$ using $\rho^2_k=h_{c}^{2}({{f}_{k}}) / 12{{\pi }^{2}}f_{k}^{3}$.

For our simulations, a total of 200 samples are drawn from \texttt{holodeck}'s multivariate distribution on $\bm{\rho}$ according to \autoref{pois}, based on the single choice of $\bm{\theta}$ in \autoref{astro-chosen}. These 200 samples, along with a unique random draw from the multivariate Gaussian distribution of \autoref{a_gwb_draw} for each of the 200 GWB realizations, represent the only variation across the 200 GWB realizations within each simulated data type. \autoref{sim_pos} shows the entire $\bm{\rho}$ distribution as well as the chosen 200 randomly-drawn samples from it. The samples are indicated with gray straight lines. As shown in the figure, the sample size of 200 is sufficiently large to capture key variations in the underlying distribution, including its tails.

\end{itemize}
 \section{\label{sec: viability}Analysis of the Population Model}
We aim to quantify the sensitivity of the GWB-induced timing-residual PSD to different SMBHB population parameters. To do so, we analyze each of the 200 synthetic data realizations of \textsc{A4cast} and \textsc{SimGWB}\footnote{For both data sets, the white noise level is fixed to what is injected while performing the analyses. Thus, the matrix $D$ of \autoref{eq:likelihood} is known \textit{a priori}.}, following the Bayesian framework described in \S\ref{sec: PTA}, and as implemented in the software package \texttt{pandora} \citep{Laal_PANDORA_2025}, which uses the MCMC package \texttt{PTMCMC} \citep{PTMCMC} for sampling. The focus of our analyses is on the marginal posterior distributions of the SMBHB parameters; however, in \autoref{app:joint}, we briefly and qualitatively discuss joint posterior probability distributions. 

\begin{table*}
\centering
\makebox[\linewidth][c]{%
\begin{tabular}{ccccccc}
\toprule
       & $\tau_f$ & $\psi_{0}$ & $m_{\psi,0}$ & $\mu$ & $\epsilon_{\mu}$ & $\nu_{\rm{inner}}$ \\
\midrule
\textsc{A4cast} & ${0.20}^{+0.10}_{-0.11}$    & ${0.10}^{+0.09}_{-0.05}$           &${0.20}^{+0.04}_{-0.03}$                              &${0.04}^{+0.02}_{-0.01}$    &${0.06}^{+0.03}_{-0.02}$                                 &${0.34}^{+0.15}_{-0.17}$                    \\
\textsc{SimGWB} &   ${0.35}^{+0.07}_{-0.07}$  & ${0.22}^{+0.11}_{-0.09}$            & ${0.26}^{+0.03}_{-0.02}$                              & ${0.05}^{+0.02}_{-0.01}$    & ${0.09}^{+0.07}_{-0.04}$                                 & ${0.59}^{+0.05}_{-0.07}$                  \\
\bottomrule
\end{tabular}
}
\caption{Hellinger distance values between the prior and the posterior distribution for all of the parameters within $\bm{\theta}$ for both \textsc{A4Cast} and \textsc{simGWB}. The range of the Hellinger distances is over all 200 realizations of each simulated data set. Each entry in the table represents the median of the 200 Hellinger distance values of each dataset's SMBHB population parameter, with the error bars indicating the 1-$\sigma$ equivalent uncertainty range.}
\label{table:parameters}
\end{table*}

 \begin{figure}
\includegraphics[width=\linewidth]{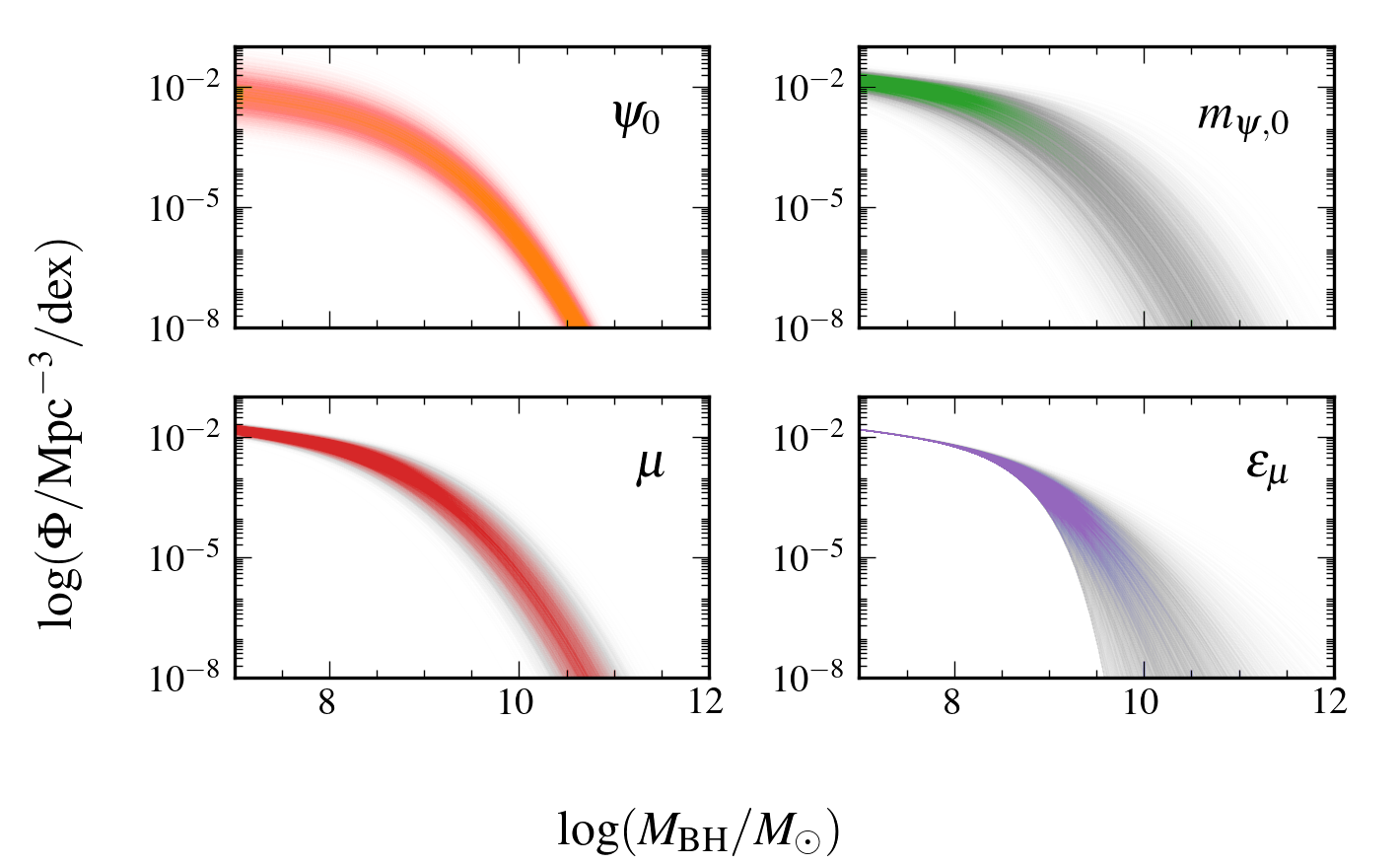}
    \caption{A series of plots showing the relationship between the black hole mass function $\Phi$ and the mass of black holes. The label of each subplot indicates the \textsc{phenom} model parameter that varies over its prior distribution samples, while the rest of the parameters are fixed to their values in \autoref{astro-chosen}. Each thin band in the plot corresponds to a unique draw from the relevant prior distribution. An arbitrary redshift of 1 is chosen for the plots. Note that $\tau_f$ and $\nu_{\rm{inner}}$ do not affect the black hole mass function; hence, they are not relevant for this plot. Compared to the other two parameters, changes in $\psi_0$ and $m_{\psi,0}$ over their prior distribution correspond to a larger change in the black hole mass function both at low and high black hole mass values.}
    \label{BHMF}
\end{figure}

 \subsection{\label{sec: viability-Bayesres}Bayesian inferences: posterior versus prior} 
 \begin{figure*}[htbp]
\centering
\subfloat{\includegraphics[width=0.49\textwidth]{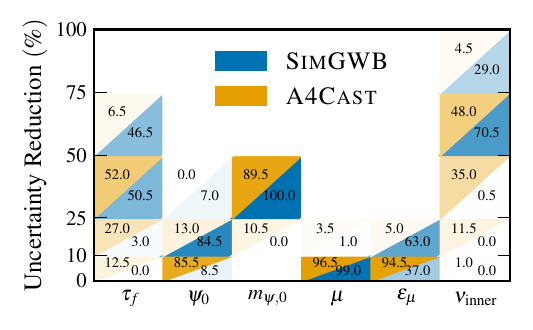}}
\subfloat{\includegraphics[width=0.49\textwidth]{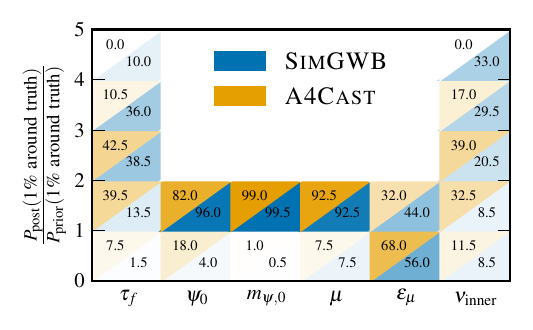}}
\caption{Heatmap plots depicting the reduction in uncertainty (68\% HDI) of the posterior relative to the prior (left panel) and the posterior-to-prior probability ratio for the small region around the actual values (values of \autoref{astro-chosen} $\pm$1\%) (right panel). The numbers within each cell indicate the percentage of total realizations that fall within a range of the estimated quantity in each panel. The two data sets are color-coded: different shades of blue (lower triangular region of each cell) correspond to \textsc{simGWB}, and different shades of orange (upper triangular region of each cell) correspond to \textsc{A4Cast}. For instance, for $\tau_f$, 50.5\% of the realizations see a reduction in uncertainty between 25 to 50\% if the inferences are done on \textsc{simGWB}, while this number is 52\% for \textsc{A4Cast} data. Overall, the figure quantifies the extent of information gain after incorporating the PTA data.}
\label{CI}
\end{figure*}
Within each panel of \autoref{postcomp}, posterior distributions resulting from the Bayesian analyses of all simulated data sets are reduced to posterior credible intervals. They are depicted as scattered points, with error bars denoting the maximum a-posteriori (MAP) and the 68-percent highest density interval (HDI) \citep{arviz_2019}, respectively. The prior on each SMBHB parameter (\autoref{astro-prior}) is shown in the small rectangular sub-panel to the right of the main panel. Furthermore, the markers are placed in ascending order from left to right based on the size of the error bar of the $\tau_f$ posterior distributions\footnote{This choice is entirely arbitrary and is meant to assist with the readability of \autoref{postcomp}.}. In the following, we highlight a few key points.

 \subsubsection{Constrainability}
Not all parameters are equally constrainable. For $\mu$, only uninformative posterior distributions are observed over all realizations. This is evidenced by the low Hellinger distance\footnote{Hellinger distance measures the similarity between two probability distributions, bounded in the range $[0,1]$; the lower this value, the more consistent are the two distributions. More precisely, for two discrete probability distributions $P = (p_1, p_2, \ldots , p_r)$ and $Q = (q_1, q_2, \ldots , q_r)$, the Hellinger distance $H$ between $P$ and $Q$ is defined as 
\begin{equation}
    H = \frac{1}{\sqrt{2}} \sqrt{\sum\limits_{i=1}^{r}{{{\left( \sqrt{{{p}_{i}}}-\sqrt{{{q}_{i}}} \right)}^{2}}}}.
\end{equation} 
An intuition for Hellinger distances was given by \citet{fitting}, where two equal-variance Gaussian distributions whose means are offset by $1(2)$-$\sigma$ have Hellinger distances of $0.34(0.63)$.} between the posteriors of individual realizations and the prior distribution, regardless of the type of simulation; see \autoref{table:parameters}. This indicates that there are intrinsic limitations to how well PTAs can constrain $\mu$, as well as $\varepsilon_{\mu}$, which also shows uninformative posterior constraints. For the remaining parameters, PTA data is shown to be informative to varying degrees, either by reducing astrophysical parameter uncertainties or by pushing the MAP closer to the true value. Refer to \autoref{table:parameters} for the range (over all realizations) of computed Hellinger distances between posterior and prior distributions of the SMBHB parameters. The table quantifies the informativeness of Bayesian inference: the higher the Hellinger distance, the more informative it is.

To understand why PTA data can aid in constraining some SMBHB parameters but is more limited for others, we recall some of the discussion from \S\ref{sec:library}. The evolutionary parameters $\tau_f$ and $\nu_{\text{inner}}$ play a fundamentally different role from the other components of $\bm{\theta}$. Specifically, $\tau_f$ and $\nu_{\text{inner}}$ determine when and whether existing SMBHBs enter the PTA frequency band, whereas the remaining four parameters describe the overall black hole mass function, $\Phi$, which governs the number density of black holes as a function of mass, independent of whether they fall within the PTA band. Before performing the analyses, it is challenging to predict whether the GWB signal will be more sensitive to the parameters controlling binary evolution into the PTA band or to those affecting the black hole mass function.

Nevertheless, we can \textit{a priori} identify those parameters to which the black hole mass function, $\Phi$, is most sensitive. \autoref{BHMF} depicts how variations of $\psi_0$, $m_{\psi ,0}$, $\mu$, and $\varepsilon_{\mu}$ within their respective prior distributions impact the black hole mass function. The evolution parameters $\tau_f$ and $\nu_{\text{inner}}$ do not affect the black hole mass function, as the black hole mass function controls the number density of all black holes, not necessarily the ones that contribute to the GWB detected by PTAs. In each panel, the \textsc{phenom} parameter indicated by the label is varied according to its prior distribution while the other three parameters are fixed to their fiducial values in \autoref{astro-chosen}. As evident in \autoref{BHMF}, when compared to $\mu$ and $\varepsilon_{\mu}$, variations in $\psi_0$ and $m_{\psi,0}$ over their prior distribution correspond to a larger change in the black hole mass function at both low and high black hole mass values. For the range of black hole masses relevant to PTAs ($>10^7 M_\odot$), the black hole mass function is more sensitive to the galaxy stellar mass function parameters $\psi_0$ and $m_{\psi,0}$ than to the $M_{\text{BH}}$--$M_{\text{Bulge}}$ parameters. Thus, the greater extent of prior-dominance for $\mu$ and $\varepsilon_{\mu}$ over $\psi_0$ and $m_{\psi ,0}$ shown in \autoref{postcomp} is to be expected.

In the existing PTA literature, \citet{2016ApJ...826...11S} identifies the $M_{\text{BH}}$--$M_{\text{Bulge}}$ as having the largest effect on the mean value of the characteristic strain amplitude of the GWB. However, this does not necessarily imply that the $M_{\text{BH}}$$–$$M_{\text{Bulge}}$ parameters have the greatest influence on the full distribution of the GWB strain amplitude. In this work, we explicitly account for the entire distribution of the GWB characteristic strain generated by \texttt{holodeck} under the \textsc{phenom} model when constructing the prior probability distribution for the Bayesian inferences, thanks to the use of normalizing flows \citep{LaalAstro}. Thus, our results do not contradict those of \citet{2016ApJ...826...11S}.

 \subsubsection{Dataset regimes}
As expected, the significance of the GWB signal (via the realism of the data set) has a strong impact on the ability to constrain SMBHB population parameters. The left panel of \autoref{CI} quantifies the percentage reduction in uncertainty from the prior to the posterior distribution.
For each dataset's realization, the percent difference between the 68\% HDIs of the posterior and prior distributions is computed using
\begin{equation}
    R=100\left[ 1-\frac{ {{u}_{\text{pos}}}-{{l}_{\text{pos}}} }{{{u}_{\text{prior}}}-{{l}_{\text{prior}}}} \right],
\end{equation}
where $R$ is the percentage reduction in uncertainty; $u_{\text{pos}}$ and $l_{\text{pos}}$ are the upper and lower bounds of the posterior 68\% HDI; and $u_{\text{prior}}$ and $l_{\text{prior}}$ are the corresponding bounds of the prior.
The values of $R$ are grouped into five unequal bins: $[0,10)$, $[10,25)$, $[25,50)$, $[50,75)$, and $[75,100]$. Each cell in the figure shows the percentage of realizations falling into each bin. As the figure shows, \textsc{A4cast}'s $\bm{\theta}$ posterior distributions are generally less constrained than those from \textsc{SimGWB}. For context, the \textsc{A4cast}'s 68-percent HDI is, on average (over all 200 realizations), about 7 times larger than \textsc{simGWB}'s 68-percent HDI for the GWB spectrum parameter $\rho_1$. For \textsc{A4cast}, there is a greater than 50\% chance (across realizations) of reducing the uncertainty in $\nu_{\text{inner}}$ by $>$50\%, and the uncertainty in $\tau_f$ by between 25--75\%. Moreover, there is an 88\% chance of reducing the uncertainty in $m_{\psi,0}$ by 25--50\%.

These results are particularly relevant for current PTAs, which are more consistent with \textsc{A4cast} than they are with \textsc{simGWB}, since the latter is highly idealized and akin to a far-future signal significance scenario. In addition to the $M_{\text{BH}}$--$M_{\text{Bulge}}$ parameters, the power of PTAs to constrain the GSMF normalization parameter, $\psi_0$, is minimal for realistic data sets. On the other hand, there are strong prospects for PTAs to constrain the typical total lifetime of SMBHBs, $\tau_f$, the small-separation hardening rate, $\nu_{\text{inner}}$, and the characteristic mass normalization of the GSMF, $m_{\psi,0}$.

 \subsubsection{Change in belief about the true values of the SMBHB parameters}
A posterior distribution reflects how our belief about a model parameter changes after incorporating data. Naturally, we hope this updated belief yields a distribution that is more consistent with the parameter’s true value than the prior distribution; this serves as a measure of the success of a Bayesian inference. To quantify this success in our analyses of simulated data, we define a small region around the true parameter values (i.e., values of \autoref{astro-chosen} $\pm$ their 1\%) as the part of the parameter space within which we compute the probability mass\footnote{The probability mass is given by the integral of a probability density function over a region.}. The ratio of the posterior probability mass to the prior probability mass over the defined small region, referred to as ``posterior-to-prior ratio'' for brevity, quantifies the change in belief regarding the presence of the \autoref{astro-chosen} values in the simulated PTA data.

The right panel of \autoref{CI} shows the distribution of these ratios. This panel is constructed in the same way as the left panel, except that the quantity of interest is now the posterior-to-prior ratio. Evident from the right panel of \autoref{CI} and except for $\varepsilon_\mu$, the posterior distributions assign higher probability to the small region around the actual values than the prior distributions do. For example, for the \textsc{simGWB} data set and for the parameter $\nu_{\text{inner}}$, 33\% of the realizations show an increase in belief between four and five times. Although setting a threshold for the posterior-to-prior probability ratio is subjective, we consider any ratio above 2 to be a success, between 1 and 2 to be a negligible improvement, and below 1 to be a failure. Based on this criterion, the parameter recovery of the two parameters that are otherwise not easily constrainable using electromagnetic observations (i.e., $\tau_f$ and $\nu_{\text{inner}}$) is a success for both types of data sets, as more than 50\% of the realizations show a posterior-to-prior probability ratio higher than 2. The same cannot be said of the remaining four parameters; however, since they are already strongly constrained by electromagnetic observations, this is not a cause for concern. 

As shown in \citep{LaalAstro}, fixing the prior-dominated parameters to their actual values can improve parameter recovery for the remaining parameters. However, we acknowledge that our choice of values in \autoref{astro-chosen} could significantly affect the recovery of the actual values, and further work is needed to understand this effect. In other words, we do not know which to blame the most for the unsuccessful recovery of the actual values of all the SMBHB population parameters: the details of the construction of \textsc{phenom}, specific choices made while creating the simulated data sets, or the statistical properties (e.g., variance) of the GWB signal inferred from PTAs.

 \subsection{\label{sec: viability-Forestres}Sensitivity analysis using random forest regression}
We wish to understand the extent to which our findings on the constraining influence of PTA data on SMBHB astrophysical hyperparameters are general, or dependent on the specific value of $\bm{\theta}$. However, generating and performing a comprehensive simulation analysis using every possible value of $\bm{\theta}$ is computationally prohibitive. Therefore, instead, we employ \emph{random forest regression} \citep{random_forest}.

Random forest regression is a machine learning technique that uses \emph{decision trees} to predict a specific scalar output given any input. Relevant to the interests of this work, the structure of an established random forest regressor can be used to examine the sensitivity of the output to changes in the inputs. For instance, we can study how each parameter in $\bm{\theta}$ correlates with changes in the GWB-induced timing residual PSD, $\bm{\rho}$, at each frequency bin.

 \begin{figure}[htbp]
\includegraphics[width=\linewidth]{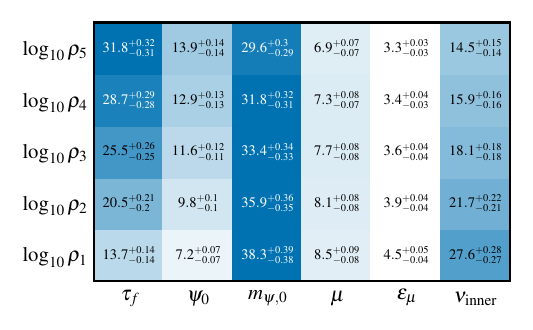}
    \caption{A heatmap plot depicting the range of random forest regression scores for each parameter at each frequency bin. The different shades of blue provide a visual clue for the score: the darker the blue, the larger the score. Completely independent decision trees are constructed for each frequency bin; thus, scores are not comparable across frequency bins. The figure indicates that $\mu$ and $\varepsilon_{\mu}$ have negligible influence on the changes of in the amplitude of the GWB-induced timing residual PSD, in contrast to the strong impact of $m_{\psi,0}$, $\nu_{\rm{inner}}$, and $\tau_f$.}
    \label{RFR}
\end{figure}

\autoref{RFR} shows the range of the feature-importance scores for each SMBHB parameter. The scores quantify the correlation of each free parameter of \textsc{phenom} with changes to each $\log_{10}{\rho}$ PSD parameter. Note that the scores are calibrated in terms of percentages, so that the sum of scores across SMBHB parameters is 100 for a given $\log_{10}{\rho}$ (i.e., scores sum to 100 in each row of \autoref{RFR}). Furthermore, one cannot compare scores across frequency bins, as random forest regression is only applicable to scalar outputs; hence, each frequency bin requires its own independently constructed random forest regressor. Based on \autoref{RFR}, the two parameters $\mu$ and $\varepsilon_{\mu}$ have the lowest scores ($<10\%$) across all frequency bins, signifying that the PSD has negligible sensitivity to changes in these parameters. On the other hand, parameters ${{m}_{\psi,0}}$, $\nu_{\rm{inner}}$, and $\tau_f$ have the highest scores, which implies a significant level of sensitivity of the PSD to variations in these three parameters. 

These results generalize our earlier claim about the limited influence of $M_{\text{BH}}$--$M_{\text{Bulge}}$ relation parameters on the GWB compared to GSMF parameters shown in \autoref{BHMF}. Additionally, the scores confirm the findings of the earlier Bayesian inference analyses. 
As concluded in \S\ref{sec: viability-Bayesres}, and now reaffirmed using random forest regression, only three out of the six free parameters of \textsc{phenom} are meaningfully constrainable. These parameters are ${{m}_{\psi,0}}$, $\nu_{\rm{inner}}$, and $\tau_f$.
 
\section{\label{sec: Conclusion}Conclusion \& Implications}
This work was motivated by our hypothesis that many detailed astrophysical (or potentially even cosmological) models for the source of the GWB are underconstrained by PTA data, as they contain too many parameters, most of which overlap in their effects on the GWB signal. To test this hypothesis, we examined the viability of an extensively used model for SMBHB population and evolution for astrophysical inferences, named \textsc{phenom} \citep{15yrastro, Chen2019}, by performing Bayesian inference on simulated data with varying degrees of realism. To further contextualize the results of inferences, we estimated feature-importance scores for \textsc{phenom} parameters using random forest regression and analyzed the black hole mass function's sensitivity to changes in relevant \textsc{phenom} parameters. For the Bayesian analyses, we treated the SMBHB population and evolution parameters as part of a typical PTA data model, with GWB PSD parameters simply treated as intermediate quantities that bridge to the astrophysical parameters of interest.

In this work, we have outlined methods for performing rigorous statistical assessments of a model for the SMBHB population, while accounting for the complications of modeling a GWB using typical PTA modeling techniques. Our framework can be easily adopted and built upon in the future when considering different models for the isotropic GWB and SMBHB populations \citep{evolvingMM}, combining the anisotropic nature of GWB with galaxy distribution through cross-correlation analyses \citep{Sah:2025uuk}, as well as cosmological GWB models. 

Using our statistical framework, our results show a prior-mimicking tendency in the inference of SMBHB population parameters with existing strong non-PTA astrophysical constraints. In this case, the results are shaped more by our prior choice than by the PTA data, regardless of data quality. More explicitly, the parameters that control the $M_{\text{BH}}$--$M_{\text{Bulge}}$ parameters within \textsc{phenom} are always prior-dominated, no matter how realistic the PTA data is. On the other hand, there are prospects for information gain for parameters that control the total lifetime of the SMBHB population, the hardening rate, and the characteristic mass normalization of the GSMF.

Our results highlight the importance of a multimessenger approach to studying the SMBHB population. As shown in this work, a PTA-inferred GWB cannot further constrain parameters that are already well constrained electromagnetically, such as the $M_{\text{BH}}$--$M_{\text{Bulge}}$ relation parameters. On the other hand, PTAs are excellent probes of SMBHB parameters that are unconstrained by electromagnetic observations, such as the total lifetime of the SMBHB population or the hardening rate. Hence, population multi-messenger probes of SMBHBs with PTA data will help to illuminate the dynamical processes governing SMBHB evolution at small scales, while also informing the requirements of larger-scale demographic studies from electromagnetic data.

\subsection{\label{sec:software}Software}
We took advantage of the functionalities provided by \texttt{holodeck} \citep{15yrastro}, \texttt{PTMCMC sampler} \citep{PTMCMC}, \texttt{pytorch} \citep{pytorch}, \texttt{zuko} \citep{rozet2022zuko}, \texttt{jax} \citep{jax2018github}, \texttt{scikit-learn} \citep{scikit-learn}, and \texttt{pandora} \citep{Laal_PANDORA_2025}. Python package \texttt{matplotlib} \citep{plt} was used for generating the figures in this paper.
\begin{acknowledgments}
The NANOGrav NSF Physics Frontier Center supported our work awards \#2020265 and \#1430284. S.R.T acknowledges support from NSF AST-2307719, and an NSF CAREER \#2146016. S.R.T also acknowledges a Chancellor's Faculty Fellowship from Vanderbilt University. N.L is supported by the Vanderbilt Initiative in Data Intensive Astrophysics (VIDA) Fellowship. This work was conducted in part using the resources of the Advanced Computing Center for Research and Education (ACCRE) at Vanderbilt University, Nashville, TN. This work was performed in part at the Aspen Center for Physics, which is supported by National Science Foundation grant PHY-2210452.
\end{acknowledgments}

\appendix
\section{\label{app:joint}Joint constraints}
\begin{figure}[t]
\includegraphics[width=\linewidth]{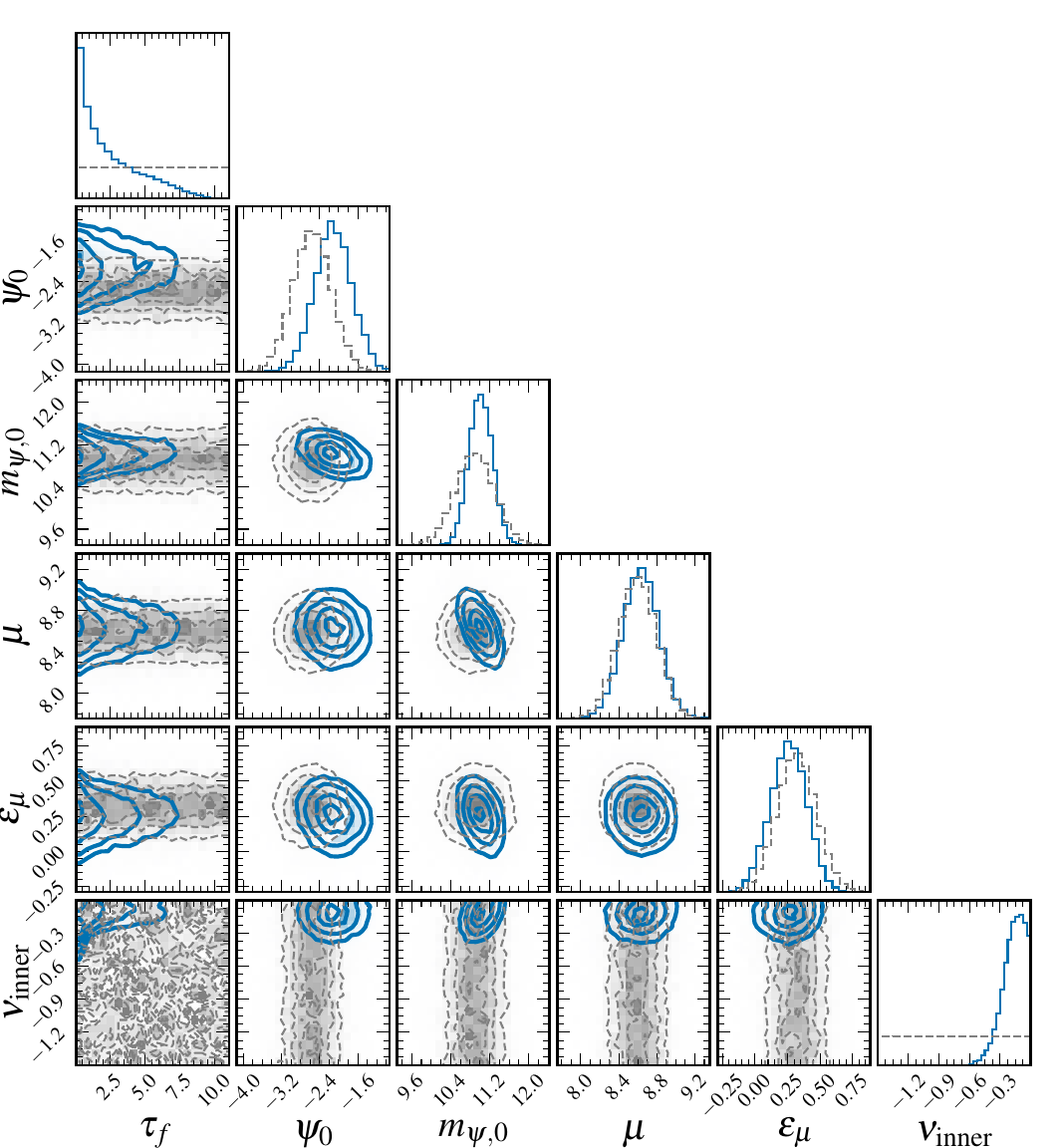}
    \caption{A corner plot depicting the 1- and 2-dimensional histograms of the SMBHB population parameters, before (i.e., prior) and after (i.e., posterior), incorporating a random realization of the \textsc{SimGWB} data set. The posterior distribution is shown in blue color using solid curves, while the prior distribution is shown in gray color and dashed curves. The plot highlights the information gain from PTAs in constraining the joint distribution of SMBHB population parameters: the prior distribution lacks information about which combinations of SMBHB parameters are allowed, whereas the posteriors provide this information.}
    \label{corner_multivar}
\end{figure}

Beyond the individual marginalized posterior distributions, Bayesian inferences provide information about which combination of SMBHB parameters best describes the PTA data. \autoref{corner_multivar} depicts such combinations for a random realization from the \textsc{SimGWB} data set. In particular, a comparison between the posterior (solid blue curves) and prior (dashed gray curves) contours shows the value of this information. Even though $M_{\text{BH}}$--$M_{\text{Bulge}}$ parameters have prior-dominated marginalized posterior distributions, their joint posterior distributions with other parameters are completely different from the joint prior distributions.

\autoref{corner_multivar} shows that PTAs can also constrain parameter combinations, revealing directions in parameter space where the data provide information despite the individual marginals being prior-dominated. This is valuable information as it reveals covariances between the SMBHB parameters that are absent in our prior knowledge. For instance, the tilted elliptical shape of $m_{\psi ,0}$ to $\mu$ contour plot signifies a negative covariance between the two parameters, such that as $\mu$ increases $m_{\psi ,0}$ decreases accordingly. Performing a quantitative analysis of the extent of covariance between the parameters across all realizations is outside the scope of this paper. We simply highlight an aspect of the Bayesian inference that reveals the correlations encoded in the joint posterior distributions.

\FloatBarrier
\bibliography{sample631}{}
\bibliographystyle{aasjournal}
\end{document}